\documentclass[manuscript]{aastex}

\usepackage{graphicx}
\usepackage{amssymb}
\usepackage{natbib}
\usepackage{amsmath}
\usepackage{url}

\slugcomment{}

\shorttitle{Plasma diagnostics of an EIT wave observed by Hinode/EIS and SDO/AIA}
\shortauthors{Veronig et al.}

\begin{document}

\title{Plasma diagnostics of an EIT wave observed by Hinode/EIS and SDO/AIA}

\author{A. M. Veronig}
\affil{Institute of Physics, University of Graz,
    Universit\"atsplatz 5, A-8010 Graz, Austria}
    \email{astrid.veronig@uni-graz.at}

\author{P. G\"om\"ory}
\affil{Astronomical Institute, Slovak Academy of Sciences, SK-05960 Tatransk\'a Lomnica, Slovakia;
Kanzelh\"ohe Observatory/Institute of Physics, University of Graz, A-9521 Treffen, Austria
}

\author{I. W. Kienreich}
\affil{Institute of Physics, University of Graz,
    Universit\"atsplatz 5, A-8010 Graz, Austria}

\author{N. Muhr}
\affil{Institute of Physics, University of Graz,
    Universit\"atsplatz 5, A-8010 Graz, Austria}
    
\author{B. Vr\v{s}nak}
\affil{Hvar Observatory, Faculty of Geodesy, Ka\v{c}i\'ceva 26, 1000 Zagreb, Croatia}

\author{M. Temmer}
\affil{Institute of Physics, University of Graz,
    Universit\"atsplatz 5, A-8010 Graz, Austria}

\author{H. P. Warren}
\affil{Space Science Division, Naval Research Laboratory, Washington, DC 20375, USA}

\begin{abstract}
We present plasma diagnostics of an EIT wave observed with high cadence in Hinode/EIS sit-and-stare spectroscopy and SDO/AIA imagery obtained during the HOP-180 observing campaign 
on 2011 February 16. At the propagating EIT wave front, we observe downward plasma flows 
in the EIS Fe\,{\sc xii}, Fe\,{\sc xiii}, and Fe\,{\sc xvi} spectral lines (log~$T \approx 6.1-6.4)$
with line-of-sight (LOS) velocities up to $20$~km~s$^{-1}$.
These red-shifts are followed by blue-shifts with upward velocities up to $-5$~km~s$^{-1}$ indicating 
relaxation of the plasma behind the wave front.
During the wave evolution, the downward velocity pulse steepens from a 
few km~s$^{-1}$ up to 20~km~s$^{-1}$ and subsequently decays, 
correlated with the relative changes of the line intensities.
The expected increase of the plasma densities at the EIT wave front estimated from the observed intensity increase 
lies within the noise level of our density diagnostics from EIS Fe\,{\sc xiii} 202/203~{\AA} line ratios. 
No significant LOS plasma motions are observed in the He\,{\sc ii} line, 
suggesting that the wave pulse was not strong enough to perturb the underlying chromosphere. 
This is consistent with the finding that no H$\alpha$ Moreton wave was associated with the event.
The EIT wave propagating along the EIS slit reveals a strong deceleration of 
$a \approx -540$~m~s$^{-2}$ and a start velocity of $v_0 \approx 590$~km~s$^{-1}$. 
These findings are consistent with the passage of a coronal fast-mode MHD wave, pushing the plasma downward and compressing
it at the coronal base.
\end{abstract}


\keywords{Sun: corona --- Sun: coronal mass ejections (CMEs) --- Sun: flares}

\section{Introduction}

Large-scale disturbances propagating through the solar corona have been first imaged by the 
Extreme-Ultraviolet Imaging Telescope (EIT) onboard the Solar and Heliospheric Observatory (SOHO) 
about 15 years ago \citep{thompson98}, and colloquially termed ``EIT waves''. 
EIT waves have been initially interpreted as the coronal counterparts of Moreton waves, 
in accordance with the fast-mode coronal MHD wave model developed by \cite{uchida68}.
However, due to statistical differences derived in the propagation velocities of Moreton waves, which are of the order of
1000 km~s$^{-1}$, and EIT waves, which lie mostly in the range 200--400 km~s$^{-1}$
\citep{klassen00,thompson09}, this interpretation was questioned and alternative models were developed.
Interpretations of EIT waves can be subdivided into wave versus non-wave models.
In the non-wave models, EIT waves are explained by the large-scale coronal restructuring
due to the erupting CME causing the observed emission enhancements either due to plasma compression, heating or localized energy release
\citep[e.g.][]{delanee99,chen02,attrill07}. 

Coronal wave studies with the EIT instrument were limited by its 12~min observing cadence. This situation drastically improved by the launch of the 
Solar Terrestrial Relations Observatory (STEREO) in October 2006, which has led to significant progress in the understanding of these intriguing events. The multi-point observing platforms of the twin STEREO satellites with their Extreme Ultraviolet Imagers \citep[EUVI;][]{howard08} 
have made it possible to gain insight into the three-dimensional structure and evolution of EUV waves \citep{patsourakos09b,kienreich09,ma09,temmer11}, including events where the full three-dimensional coronal wave dome was observed \citep{veronig10}. 
In addition, the high cadence (2.5 min), large field-of-view, and simultaneous observations from two vantage points facilitated the first detailed studies 
on the wave dynamics and its relation to the associated CME evolution \citep{patsourakos09,kienreich09,temmer11}. Recent STEREO/EUVI findings include 
observations of EIT waves, which undergo a significant deceleration during their propagation, accompanied by a decay of the perturbation amplitude and broadening of the wave front 
\citep{veronig08,veronig10,long11,muhr11}. A set of four homologous EIT waves studied in \cite{kienreich11} revealed a distinct correlation between the wave speed and the magnetosonic Mach number. 
These properties are consistent with the behavior of a (weak) fast-mode shock wave.
It is important to note that these recent observational improvements also led to significant advancement of MHD simulations of EIT waves \cite[e.g.][]{cohen09,downs11}. 
For recent reviews see \cite{wills10}, \cite{gallagher10} and \cite{zhukov11}.

Since 2010, the Atmospheric Imaging Assembly \citep[AIA;][]{lemen11} onboard the Solar Dynamics Observatory (SDO) delivers ultra-high cadence (12~s) multiwavelength imaging of the solar atmosphere. 
First studies of EIT waves with AIA revealed unexpected fine structures within the 
diffuse wave front \citep{liu11}. 
Due to the high sensitivity of AIA, also more EIT waves 
with a three-dimensional dome are reported \citep[e.g.][]{kozarev11}. 
What is still missing, however, is plasma diagnostics of EIT waves, i.e. density, temperature and plasma flows at the wave front. Such information cannot be obtained with EUV imagers but needs spectroscopic observations. 
\cite{harra03} studied an EIT wave with the Coronal Diagnostics Spectrometer (CDS) onboard SOHO, without detection of
significant line-of-sight (LOS) velocities at the wave front (i.e. $v_{\rm LOS} \lesssim 10$~km~s$^{-1}$). 
Two studies related to EIT waves were performed with the Extreme-Ultraviolet Imaging Spectrometer \citep[EIS;][]{culhane07} onboard Hinode \citep{asai08,chen10}
but the EIS observing mode was not suitable to study the plasma characteristics and motions at the EIT wave front.
In this letter, we present EIS plasma diagnostics  
of the EIT wave of 2011 February 16 based on a unique data set obtained during our Hinode Observing Plan 
HOP-180\footnote{http://www.isas.jaxa.jp/home/solar/hinode\underline{ }op/hop.php?hop=0180}, where we combined 
high-cadence sit-and-stare EIS spectroscopy with high-cadence imaging by AIA. 
EIS LOS plasma motions and line widths for the event under study have been presented in \cite{harra11}, whereas here we concentrate on the 
relation between the different plasma parameters (including plasma densities and flows) at the wave front and their evolution.

\section{Data}

\begin{figure}[p]
\resizebox{16.5cm}{!}{\includegraphics{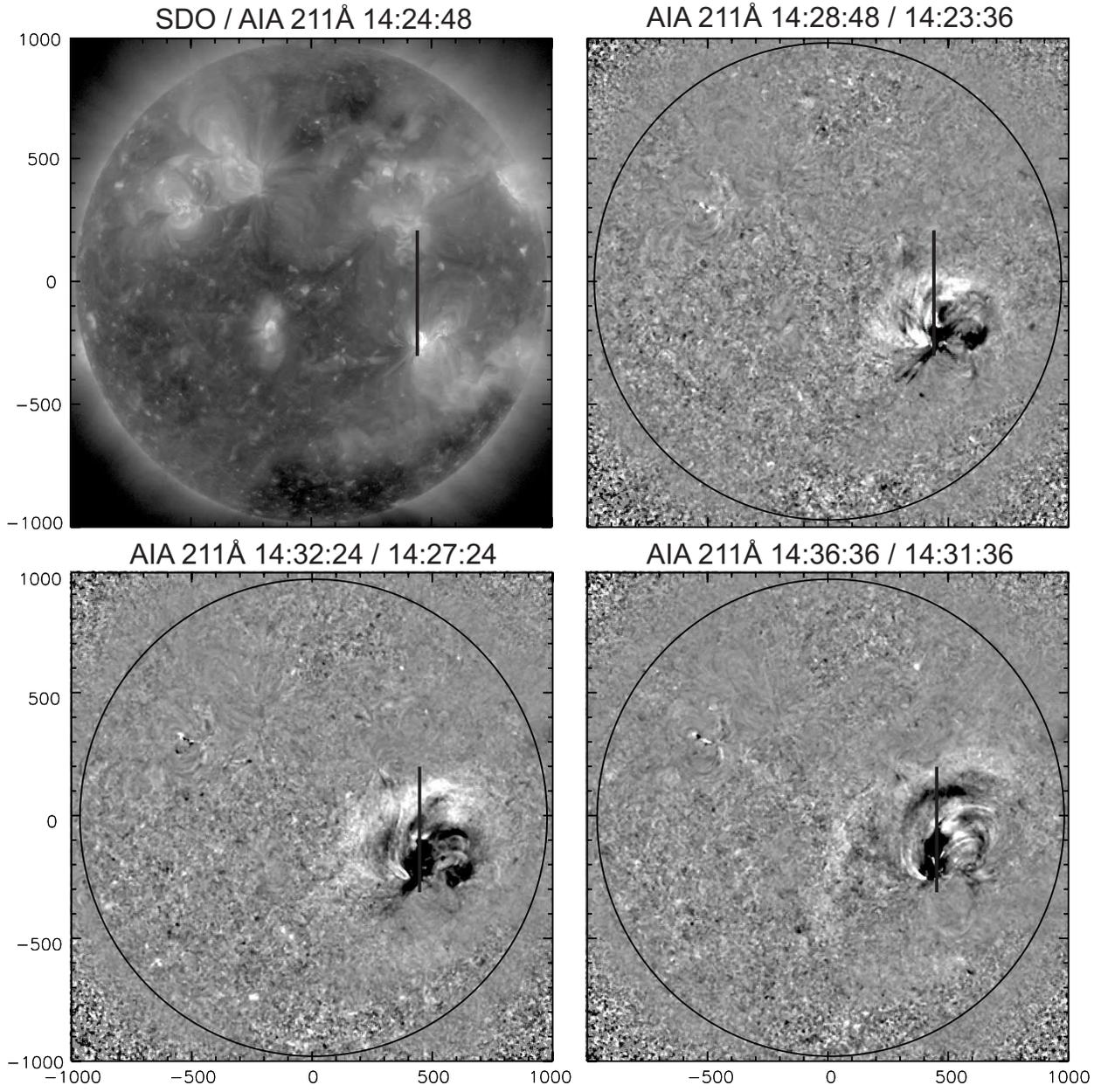}}
\caption{Sequence of SDO/AIA 211~{\AA} direct (top left) and running ratio images. The vertical line indicates the position of the Hinode/EIS 
spectrometer slit. The accompanying movie no.~1 shows the EIT wave evolution in SDO/AIA 211~{\AA} direct and running ratio images.
Movie no.~2 shows the same in the SDO/AIA 193~{\AA} filter.
} \label{fig1}
\end{figure}

Hinode/EIS is a two-channel, normal-incidence EUV spectrometer observing in the wavelength ranges 170--210~{\AA} and 250--290~{\AA} \citep{culhane07}. 
The high EIS spectral resolution allows Doppler velocity measurements of plasma flows better than $\pm$5~km~s$^{-1}$. 
In order to study the plasma characteristics at propagating EIT wave fronts, we defined a dynamic EIS observing programme, which combines
high-cadence Hinode/EIS spectroscopy and SDO/AIA imaging. This programme was accepted as HOP-180
and realized during 2011 February 11--17. In HOP-180 we performed high-cadence EIS spectroscopic sit-and-stare observations (45~s exposure + 4~s readout time), placing the spectrometer slit (width: 1$''$, lengths: 512$''$) on the border of an active region, in order to follow the evolution of EIT waves along the EIS slit with a 
spatial sampling of 1$''$ per pixel. We selected 11 EIS spectral lines over the temperature range log~$T = 4.7$ to 6.7, 
including line pairs from the same ion for density diagnostics. 
On 2011 February 16 EIS observed an EIT wave propagating along its slit (see Fig.~\ref{fig1}), on which we concentrate in this study.

Basic EIS photometric correction and correction of the orbital motion of the satellite were applied using the
eis\underline{ }prep.pro and eis\underline{ }wave\underline{ }corr.pro routines, before the spectral profiles 
were fitted by a single Gaussian function with a linear background to obtain the
spectral intensities, integrated intensities, background intensities, Doppler shifts and spectral widths.
The zero reference of the Doppler shifts were calculated as the average value of the Doppler shifts
from quiet Sun regions. From the Fe\,{\sc xiii} 202/203~{\AA} line pair we derived the coronal electron densities from 
the theoretical variation of the line ratio with density using the CHIANTI database version 6.0.1.\ \citep{dere97,dere09}.
Once the theoretical ratio was known, the final density maps were calculated using the EIS routine eis$\underline{~}$density.pro.

The EIT wave under study was also observed by SDO/AIA, which carries four (E)UV telescopes providing 
full-Sun images in ten different wavelengths at a cadence as high as 12~s and spatial resolution of 1.5$''$ (with 0.6$''$ pixels). In this study, we use in particular the AIA 211~{\AA} (log~$T = 6.3$) and 193~{\AA} (log~$T = 6.1$) filters, in which the EIT wave signal was highest.

\section{Results}

Figure~1 shows a sequence of SDO/AIA 211~{\AA} direct and running ratio images together with the location of the Hinode/EIS
spectrometer slit.  The EIT wave was launched in association with the M2 flare/CME event from AR~11158, and revealed a global propagation mainly towards  
the Northern hemisphere. In particular, a distinct segment of the wave propagated along the EIS slit northward towards AR~11159 (see also movies no.~1 and~2).

\begin{figure}[p]
\resizebox{16cm}{!}{\includegraphics{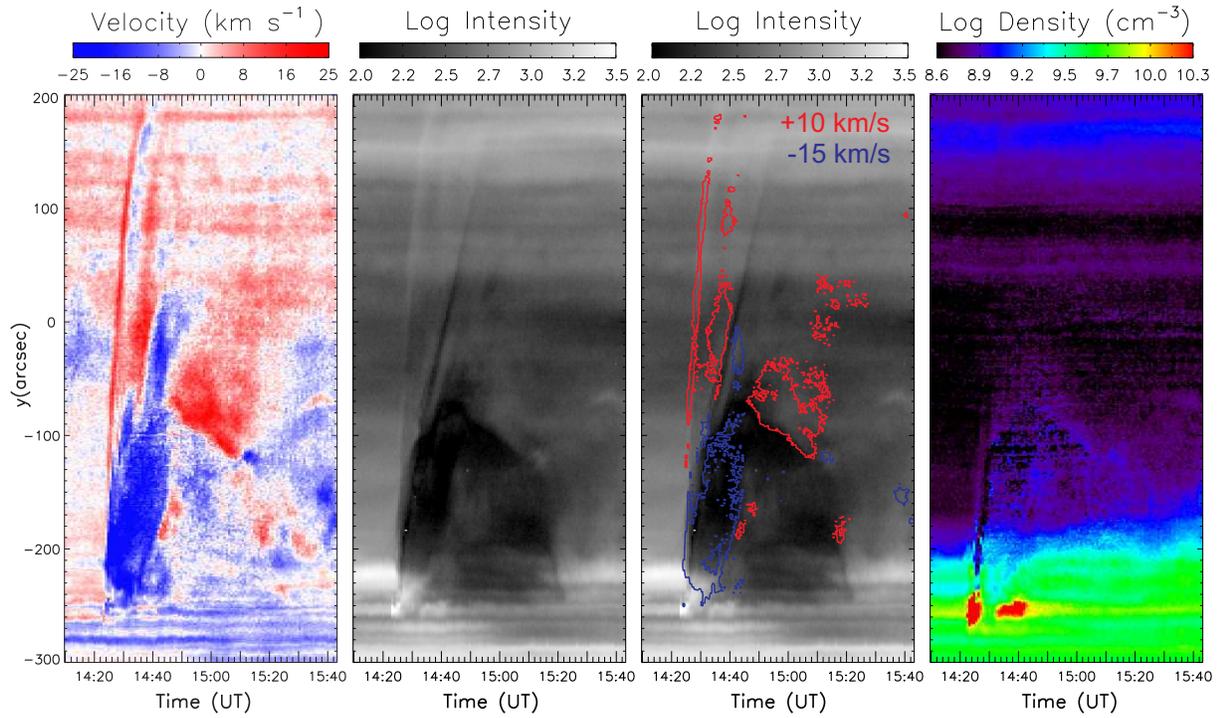}}
\caption{Stack plots showing the Fe\,{\sc xiii} 202~{\AA} evolution along the EIS slit during the time range 14:10 to 15:42 UT:
a) velocities, b) intensities, c) intensities together with velocity contours at $-15$~km~s$^{-1}$ (blue) and $+10$~km~s$^{-1}$ (red), d) densities derived from the
Fe\,{\sc xiii}~202/203~{\AA} line pair.
} \label{fig2}
\end{figure}

Figure~\ref{fig2} shows stack plots of LOS velocities and intensities derived from the EIS Fe\,{\sc xiii}~202~{\AA} spectra. 
During about 14:23 to 14:38~UT we observe a narrow lane of red-shifts indicating LOS velocities up to 20~km~s$^{-1}$ (Fig.~\ref{fig2}a) correlated with a bright lane 
in the Fe~{\sc xiii} intensities (Fig.~\ref{fig2}b,c) with enhancements up to about 25\% above the pre-event level. 
Co-aligned SDO/AIA images confirm that these signatures are due to the EIT wave front passing the EIS slit (see movie~1).
The peaks of the EIS intensity enhancements at the EIT wave front tend to occur delayed by one time step
($\approx$\,49~s) with regard to the EIS LOS velocity pulses, suggesting that the intensity change is a reaction to the downward push 
of the coronal plasma below the wave front. 
The brightest feature around 14:22~UT at $y \approx -255''$ is due to the associated M2 flare \citep{harra11}.

The EIT wave front visible as a narrow lane of red-shifts in the EIS LOS velocities is followed by blue-shifted pattern indicating relaxation 
of the plasma behind the wavefront, with upward velocities $|v_{\rm LOS}| < 5$~km~s$^{-1}$ (Fig.~\ref{fig2}a).
Behind this lane of blue-shifts there is a broader lane of red-shifts indicating another propagating feature with downward-directed plasma motions and 
concurrent intensity enhancements.
Since these LOS velocities are of the same order than that of the first lane of red-shifts, i.e. up to about 20~km~s$^{-1}$, they cannot be due to another swing 
of the primary EIT wave but are rather due to a second disturbance moving behind. 
In addition, there is a distinct dark lane observed in the EIS intensitygrams (Fig.~\ref{fig2}b), indicating propagation along the EIS slit with velocities of 
$\approx$$120\pm 10$~km~s$^{-1}$, but with no reflection in the LOS velocities (Fig.~\ref{fig2}a,c). SDO/AIA images reveal that this signal is due to an ejecta moving northward along the slit (see accompanying movie no. 1).

Fig.~\ref{fig2}d shows stack plots of plasma densities derived from the EIS Fe\,{\sc xiii} 202/203~{\AA} line pair. 
At the early phase of the EIT wave evolution, there is evidence of a lane of enhanced density 
between about $y= -200''$ and $-100''$, roughly co-spatial with the propagating EIT wave. 
However, careful comparison reveals that the enhanced densities are located about 2--3 time steps ($\sim$100--150~s) behind the wave front. At the positions of the density enhancements we observe decreased intensities and plasma upflows up to about $-50$~km~s$^{-1}$ (cf.~Fig.~\ref{fig2}a,b). 
Thus, we interpret this enhanced density feature as being related to the eruption {\it behind} the EIT wave and not due to plasma
compression at the wave front itself. In the range $y=+120''$ to $+180''$ there is some indication of a 
density depletion evolving with the EIT wave.

\begin{figure}[p]
\resizebox{12cm}{!}{\includegraphics{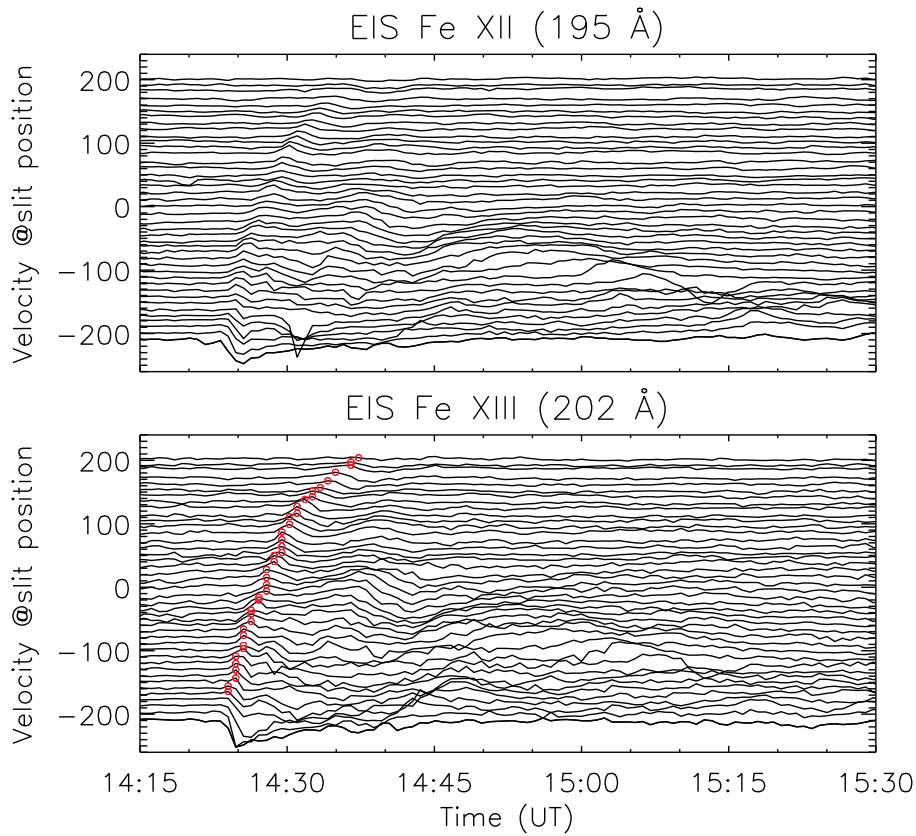}}
\caption{Velocity profiles derived from the EIS Fe\,{\sc xii} (top) and  Fe\,{\sc xiii} (bottom) spectrograms (only each 10th profile available is plotted). Each 
profile starts at the $y$-position along the slit. The circles plotted on top of the Fe\,{\sc xiii} profiles indicate the position of the 
peak LOS velocity at the wave front. These positions/values are used for the EIS wave kinematics (Fig.~\ref{fig4}a) and LOS velocity evolution (Fig.~\ref{fig4}b) as well as
for the correlation plots in Fig.~\ref{fig5}.
} \label{fig3}
\end{figure}

The distinct LOS velocity signal at the propagating EIT wave front is also well observed in the EIS Fe\,{\sc xii}~195~{\AA} spectral line (log $T = 6.11$) 
but the LOS velocities are on average 5~km~s$^{-1}$ smaller than in the Fe\,{\sc xiii}~202~{\AA} line. 
In Fe\,{\sc xvi}~262~{\AA} (log $T = 6.4$), a weak signature of the EIT wave can be observed. 
There is no significant signal of the EIT wave in the He\,{\sc ii}~256~{\AA} line (log $T = 4.7$), indicating that the upper 
chromosphere was basically unaffected by the passing EIT wavefront. This is consistent with the high-cadence H$\alpha$ observations by HASTA
(H$\alpha$ Solar Telescope for Argentinia), which show no signs of a Moreton wave. 
We also note that in the hotter spectral lines we observed no clear signal related to the EIT wave except some signature in the Ca\,{\sc xvii}~192~{\AA} line (log~$T$ = 6.7), which we attribute to a blend by Fe\,{\sc xi}.  Figure~\ref{fig3} shows cuts through the Fe\,{\sc xii} and Fe\,{\sc xiii} velocity stack plots, revealing the propagating LOS velocity pulse.
From these plots we derive the time, location and amplitude of the LOS velocity peaks used in the following to study the wave kinematics and the plasma characteristics 
with EIS. 

\begin{figure}[p]
\resizebox{10.5cm}{!}{\includegraphics{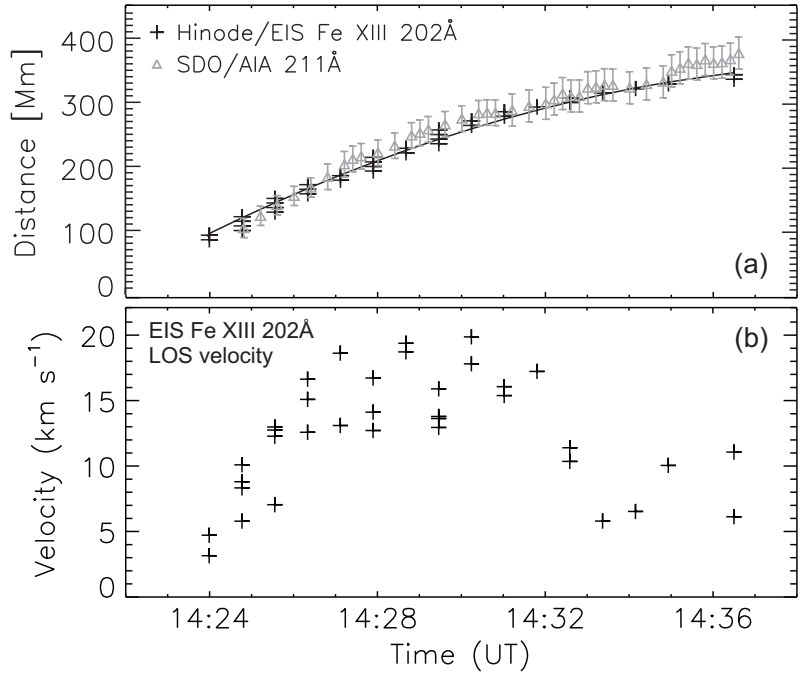}}
\caption{a) Kinematics of the EIT wave derived from the EIS Fe\,{\sc xiii} LOS velocity peaks (crosses) 
and from the wave fronts identified in SDO/AIA 211~{\AA} images (triangle with error bars). 
b) Evolution of the EIS Fe\,{\sc xiii} LOS velocity peaks derived at the EIT wave front (cf. Fig.~\ref{fig3}, bottom). 
} 
\label{fig4}
\end{figure}

In addition, we calculated the EIT wave kinematics also from SDO/AIA 211~{\AA} running ratio images. The center derived from circular 
fits to the earliest wave fronts ($x=462''\pm29''$, $y= -267''\pm 21''$) lies almost on 
the EIS slit, and we thus calculated both the AIA and EIS wave kinematics as the  
distance of the wave fronts along the spherical solar surface \citep[see][]{veronig06} to 
the pixel located at the EIS slit at $x=440''$, $y=-267''$. The resulting SDO/AIA and Hinode/EIS kinematics are shown in Fig.~\ref{fig4}a, evidencing a strong deceleration of the EIT wave. The quadratic fit to the AIA measurements gives a start velocity $v_0 = 585\pm 56$~km~s$^{-1}$ and deceleration 
$a = -675\pm 160$~m~s$^{-2}$; the mean velocity is $\bar{v} = 336\pm 15$~km~s$^{-1}$. These values are consistent with the results we obtain from calculating the EIT wave kinematics from the 
positions of the peaks of the Hinode/EIS Fe\,{\sc xiii} LOS velocities, where we find $v_0 = 587\pm 50$~km~s$^{-1}$, 
$a= -539\pm 48$~m~s$^{-2}$ and $\bar{v} = 371\pm 12$~km~s$^{-1}$.

Figure~\ref{fig4}b shows the evolution of the peak of the EIS Fe\,{\sc xiii} LOS velocity pulse at the wavefront, revealing an increase of the pulse during 14:24 UT up to 14:28 UT from 
$\sim$3--5 to 15--20~km~s$^{-1}$, and a subsequent decay during the remaining wave evolution.
We also note that the EIT wave observed in AIA imagery revealed different characteristics in different propagation directions. The fastest propagation occurred into the NW quadrant, where we find 
a start velocity $v_0 = 756\pm 56$~km~s$^{-1}$, deceleration $a=-489\pm 160$~m~s$^{-2}$ and mean velocity $\bar{v} = 576\pm 15$~km~s$^{-1}$. 

\begin{figure}[p]
\resizebox{12cm}{!}{\includegraphics{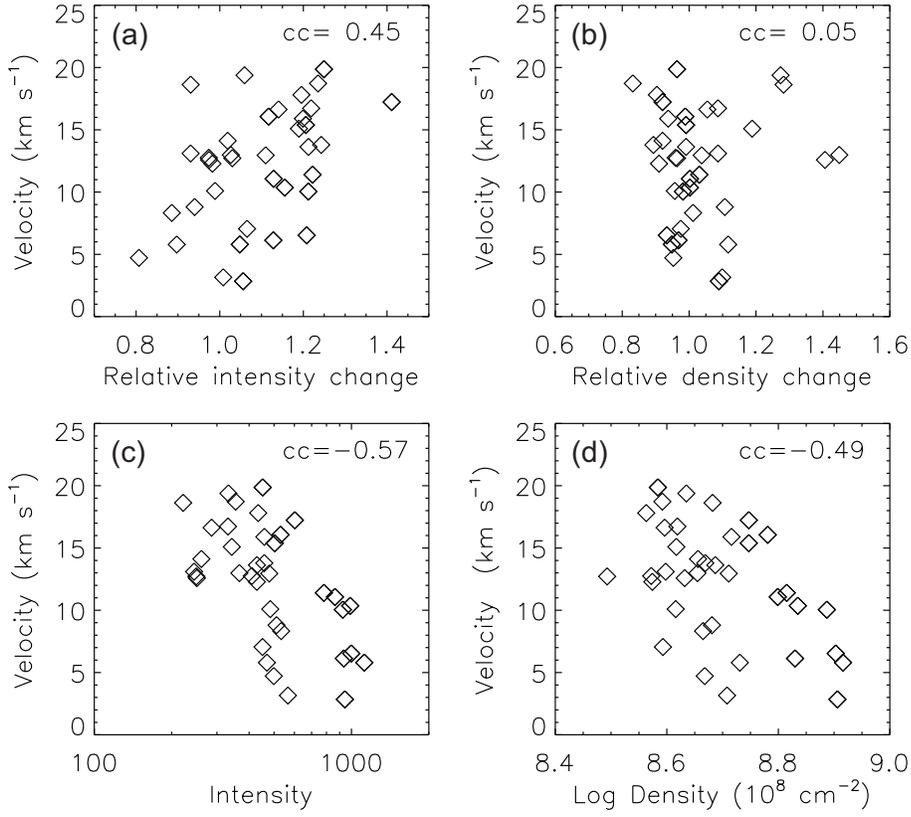}}
\caption{Correlation plots of the EIS Fe\,{\sc xiii} LOS velocity against the relative changes (top) and absolute values (bottom) 
of the intensity and density at the EIT wave front.
} 
\label{fig5}
\end{figure}

If the EIT wave under study is due to a coronal fast-mode wave pushing the plasma downward and compressing it, 
we expect a correlation between the amplitude of the velocity pulse and the change in plasma density and intensity at the wave front.
To test this hypothesis, we calculate the relative changes of these plasma parameters with respect to the 
pre-event conditions by dividing each EIS density and intensity value by the corresponding pixel prior to the EIT wave. 
The correlation plots for the EIS Fe\,{\sc xiii} parameters 
are shown in the top panels of Fig.~\ref{fig5}, where we shifted the extracted EIS intensity values by one time step (49~s) with respect to the LOS velocities to account for their delayed response.
Indeed, for the relationship LOS velocity against intensity changes we obtain a positive correlation ($cc=0.45$) indicating that large downward velocities due to the passing EIT wave front are correlated with enhanced intensities, which are usually attributed to plasma compression \cite[e.g.][]{klassen00}. However, no such correlation is found between the LOS velocity and the relative changes in density. 
This lack of correlation may be due to temperature enhancements as well as plasma compression contributing to the 
intensity changes or due to uncertainties involved in the density derivation.
The maximum changes of intensities we observe at the EIT wave front are 25\% (except for one data point; see Fig.~\ref{fig5}a). Assuming that 
the intensity enhancements at the wave front are solely due to enhanced plasma densities (and not due to changes in temperature), the maximum density increase expected is about 10\% ($I \propto n^2$ for $T=const$). This lies indeed within the noise level of our density estimates (cf.\ Fig.~\ref{fig2}d).

The bottom panels in Figure~\ref{fig5} show scatter plots of the LOS velocities against the absolute values of the EIS Fe\,{\sc xiii} intensities and densities  at the EIT wave front,
revealing a distinct anti-correlation (intensities: $cc = -0.57$, densities: $cc = -0.49$). 
The observed intensities and densities actually combine the line-of-sight integrated optically thin contributions of both the coronal ``background" plasma as well as the changes caused by the EIT wave. 
The anti-correlation obtained for the absolute values may be related to the different states of ``background" corona through which the wave propagates. 
Assuming conservation of the wave's energy flux $\rho v^2/2$, the LOS velocity caused by the passing wave front will be smaller when 
propagating through a plasma of high magnetic and gas pressure such as in an active region (appearing as horizontal stripe 
of high EUV intensity and enhanced density in Fig.~\ref{fig2}b,d) because the plasma is more inert 
than in a quiet region where the magnetic and gas pressure are small. 
This is a simplified picture, since also the wave properties themselves are expected to change during the propagation, 
in terms of intensification and decay of the wave pulse, but it can qualitatively explain the observed anti-correlations.

\section{Discussion and Conclusions}

High-cadence EIS sit-and-stare spectroscopy combined with high-cadence AIA imaging allowed us to study  
the plasma characteristics and evolution of a fast EIT wave that occurred on 2011 February 16. The fastest propagation was observed toward the NW quadrant from the source AR 11158 with a mean 
velocity of 580~km~s$^{-1}$, whereas the propagation along the EIS slit toward AR~11158 in the Northern direction revealed a mean velocity of about 370~km~s$^{-1}$.
We observe downward plasma flows in coronal spectral lines formed at temperatures in the range 1.2--2.5~MK 
co-aligned with the intensity enhancements at the propagating EIT wave front. On average the peaks of the EIS intensity 
enhancements at the EIT wave front occur delayed by $\approx 1$~min with respect to the red-shifted velocity pulses, 
suggesting that the intensity change is a reaction to the downward push of the coronal plasma below the wave front.
The peak LOS velocities of the downward plasma motions reach 
values up to 20~km~s$^{-1}$ followed by upward velocities up to $-5$~km~s$^{-1}$ indicative of relaxation of the plasma behind the passing EIT wave front. 

The downward plasma motions at the wave front reveal initial intensification and subsequent decay, correlated with the relative changes of the EIS spectral line intensities,
in line with the expectation for a fast-mode MHD wave.
However, no correlation is found between the relative density changes and the LOS velocities. This lack of correlation 
is assumed to be due to the uncertainties of the density estimates
which are of the same order as the relative density enhancements expected from the observed wave front intensities (which are $\lesssim$25\%).
To settle this issue, spectroscopic observations of EIT waves of larger amplitudes are needed. For comparison, in the EIT waves 
studied in \cite{muhr11}, relative intensity changes up to 70\% have been reported. 
In addition to the decay of the LOS velocity pulse we also observe a significant deceleration of the EIT wave during its propagation ($a \approx -540$~m~s$^{-2}$). 

Our findings are consistent with the interpretation that the EIT wave under study is a coronal fast-mode MHD wave, 
pushing the plasma downward and compressing it at the coronal base though we note that definitive density diagnostics is still missing. 
In contrast, non-wave models attributing the EIT wave to magnetic restructuring during the CME lift-off or to forced magnetic 
reconnection ahead of the CME front do not predict a downward push of the lower corona.  
In the He\,{\sc ii} line, no significant plasma motions at the EIT wave front are detected, consistent 
with the fact that no H$\alpha$ Moreton wave was associated. This finding implies that in the EIT wave under study the 
observed coronal wave pulse with LOS velocities $\lesssim$20~km~s$^{-1}$ was not strong enough to perturb the underlying ``dense'' chromosphere.

\acknowledgments AMV, PG, IWK, NM and MT gratefully acknowledge the Austrian Science Fund (FWF): P20867-N16 and V195-N16. 
PG acknowledges support by VEGA grant 2/0064/09. We thank Dr.\ Wei Liu and the anonymous referee for insightful comments on the manuscript.
Hinode is a Japanese mission developed and launched by ISAS/JAXA, with NAOJ as domestic partner and 
NASA and STFC (UK) as international partners. It is operated by these agencies in co-operation with ESA and NSC (Norway). 
We thank the SDO/AIA team for designing and operating AIA. 
CHIANTI is a collaborative project involving George Mason University,
the University of Michigan (USA) and the University of Cambridge (UK).
HASTA data are obtained at OAFA (El Leoncito, San Juan, Argentina) in the 
framework of the German-Argentinean HASTA/MICA 
Project, a collaboration of MPE, IAFE, OAFA and MPAe.


\begin{thebibliography}{35}
\expandafter\ifx\csname natexlab\endcsname\relax\def\natexlab#1{#1}\fi

\bibitem[{{Asai} {et~al.}(2008){Asai}, {Hara}, {Watanabe}, {Imada}, {Sakao},
  {Narukage}, {Culhane}, \& {Doschek}}]{asai08}
{Asai}, A., {Hara}, H., {Watanabe}, T., {Imada}, S., {Sakao}, T., {Narukage},
  N., {Culhane}, J.~L., \& {Doschek}, G.~A. 2008, \apj, 685, 622

\bibitem[{{Attrill} {et~al.}(2007){Attrill}, {Harra}, {van Driel-Gesztelyi}, \&
  {D{\'e}moulin}}]{attrill07}
{Attrill}, G.~D.~R., {Harra}, L.~K., {van Driel-Gesztelyi}, L., \&
  {D{\'e}moulin}, P. 2007, \apjl, 656, L101

\bibitem[{{Chen} {et~al.}(2010){Chen}, {Ding}, \& {Chen}}]{chen10}
{Chen}, F., {Ding}, M.~D., \& {Chen}, P.~F. 2010, \apj, 720, 1254

\bibitem[{{Chen} {et~al.}(2002){Chen}, {Wu}, {Shibata}, \& {Fang}}]{chen02}
{Chen}, P.~F., {Wu}, S.~T., {Shibata}, K., \& {Fang}, C. 2002, \apjl, 572, L99

\bibitem[{{Cohen} {et~al.}(2009){Cohen}, {Attrill}, {Manchester}, \&
  {Wills-Davey}}]{cohen09}
{Cohen}, O., {Attrill}, G.~D.~R., {Manchester}, IV, W.~B., \& {Wills-Davey},
  M.~J. 2009, \apj, 705, 587

\bibitem[{{Culhane} {et~al.}(2007){Culhane}, {Harra}, {James}, {Al-Janabi},
  {Bradley}, {Chaudry}, {Rees}, {Tandy}, {Thomas}, {Whillock}, {Winter},
  {Doschek}, {Korendyke}, {Brown}, {Myers}, {Mariska}, {Seely}, {Lang}, {Kent},
  {Shaughnessy}, {Young}, {Simnett}, {Castelli}, {Mahmoud}, {Mapson-Menard},
  {Probyn}, {Thomas}, {Davila}, {Dere}, {Windt}, {Shea}, {Hagood}, {Moye},
  {Hara}, {Watanabe}, {Matsuzaki}, {Kosugi}, {Hansteen}, \&
  {Wikstol}}]{culhane07}
{Culhane}, J.~L., {et~al.} 2007, \solphys, 243, 19

\bibitem[{{Delann{\'e}e} \& {Aulanier}(1999)}]{delanee99}
{Delann{\'e}e}, C., \& {Aulanier}, G. 1999, \solphys, 190, 107

\bibitem[{{Dere} {et~al.}(1997){Dere}, {Landi}, {Mason}, {Monsignori Fossi}, \&
  {Young}}]{dere97}
{Dere}, K.~P., {Landi}, E., {Mason}, H.~E., {Monsignori Fossi}, B.~C., \&
  {Young}, P.~R. 1997, \aaps, 125, 149

\bibitem[{{Dere} {et~al.}(2009){Dere}, {Landi}, {Young}, {Del Zanna},
  {Landini}, \& {Mason}}]{dere09}
{Dere}, K.~P., {Landi}, E., {Young}, P.~R., {Del Zanna}, G., {Landini}, M., \&
  {Mason}, H.~E. 2009, \aap, 498, 915

\bibitem[{{Downs} {et~al.}(2011){Downs}, {Roussev}, {van der Holst}, {Lugaz},
  {Sokolov}, \& {Gombosi}}]{downs11}
{Downs}, C., {Roussev}, I.~I., {van der Holst}, B., {Lugaz}, N., {Sokolov},
  I.~V., \& {Gombosi}, T.~I. 2011, \apj, 728, 2

\bibitem[{{Gallagher} \& {Long}(2010)}]{gallagher10}
{Gallagher}, P.~T., \& {Long}, D.~M. 2010, \ssr, 127

\bibitem[{{Harra} \& {Sterling}(2003)}]{harra03}
{Harra}, L.~K., \& {Sterling}, A.~C. 2003, \apj, 587, 429

\bibitem[{{Harra} {et~al.}(2011){Harra}, {Sterling}, {G{\"o}m{\"o}ry}, \&
  {Veronig}}]{harra11}
{Harra}, L.~K., {Sterling}, A.~C., {G{\"o}m{\"o}ry}, P., \& {Veronig}, A. 2011,
  \apjl, 737, L4

\bibitem[{{Howard} {et~al.}(2008){Howard}, {Moses}, {Vourlidas}, {Newmark},
  {Socker}, {Plunkett}, {Korendyke}, {Cook}, {Hurley}, {Davila}, {Thompson},
  {St Cyr}, {Mentzell}, {Mehalick}, {Lemen}, {Wuelser}, {Duncan}, {Tarbell},
  {Wolfson}, {Moore}, {Harrison}, {Waltham}, {Lang}, {Davis}, {Eyles},
  {Mapson-Menard}, {Simnett}, {Halain}, {Defise}, {Mazy}, {Rochus}, {Mercier},
  {Ravet}, {Delmotte}, {Auchere}, {Delaboudiniere}, {Bothmer}, {Deutsch},
  {Wang}, {Rich}, {Cooper}, {Stephens}, {Maahs}, {Baugh}, {McMullin}, \&
  {Carter}}]{howard08}
{Howard}, R.~A., {et~al.} 2008, Space Science Reviews, 136, 67

\bibitem[{{Kienreich} {et~al.}(2009){Kienreich}, {Temmer}, \&
  {Veronig}}]{kienreich09}
{Kienreich}, I.~W., {Temmer}, M., \& {Veronig}, A.~M. 2009, \apjl, 703, L118

\bibitem[{{Kienreich} {et~al.}(2011){Kienreich}, {Veronig}, {Muhr}, {Temmer},
  {Vr{\v s}nak}, \& {Nitta}}]{kienreich11}
{Kienreich}, I.~W., {Veronig}, A.~M., {Muhr}, N., {Temmer}, M., {Vr{\v s}nak},
  B., \& {Nitta}, N. 2011, \apjl, 727, L43

\bibitem[{{Klassen} {et~al.}(2000){Klassen}, {Aurass}, {Mann}, \&
  {Thompson}}]{klassen00}
{Klassen}, A., {Aurass}, H., {Mann}, G., \& {Thompson}, B.~J. 2000, \aaps, 141,
  357

\bibitem[{{Kozarev} {et~al.}(2011){Kozarev}, {Korreck}, {Lobzin}, {Weber}, \&
  {Schwadron}}]{kozarev11}
{Kozarev}, K.~A., {Korreck}, K.~E., {Lobzin}, V.~V., {Weber}, M.~A., \&
  {Schwadron}, N.~A. 2011, \apjl, 733, L25

\bibitem[{{Lemen} {et~al.}(2011){Lemen}, {Title}, {Akin}, {Boerner}, {Chou},
  {Drake}, {Duncan}, {Edwards}, {Friedlaender}, {Heyman}, {Hurlburt}, {Katz},
  {Kushner}, {Levay}, {Lindgren}, {Mathur}, {McFeaters}, {Mitchell}, {Rehse},
  {Schrijver}, {Springer}, {Stern}, {Tarbell}, {Wuelser}, {Wolfson}, {Yanari},
  {Bookbinder}, {Cheimets}, {Caldwell}, {Deluca}, {Gates}, {Golub}, {Park},
  {Podgorski}, {Bush}, {Scherrer}, {Gummin}, {Smith}, {Auker}, {Jerram},
  {Pool}, {Soufli}, {Windt}, {Beardsley}, {Clapp}, {Lang}, \&
  {Waltham}}]{lemen11}
{Lemen}, J.~R., {et~al.} 2011, \solphys, in press

\bibitem[{{Liu} {et~al.}(2010){Liu}, {Nitta}, {Schrijver}, {Title}, \&
  {Tarbell}}]{liu11}
{Liu}, W., {Nitta}, N.~V., {Schrijver}, C.~J., {Title}, A.~M., \& {Tarbell},
  T.~D. 2010, \apjl, 723, L53

\bibitem[{{Long} {et~al.}(2011){Long}, {Gallagher}, {McAteer}, \&
  {Bloomfield}}]{long11}
{Long}, D.~M., {Gallagher}, P.~T., {McAteer}, R.~T.~J., \& {Bloomfield}, D.~S.
  2011, \aap, 531, A42

\bibitem[{{Ma} {et~al.}(2009){Ma}, {Wills-Davey}, {Lin}, {Chen}, {Attrill},
  {Chen}, {Zhao}, {Li}, \& {Golub}}]{ma09}
{Ma}, S., {et~al.} 2009, \apj, 707, 503

{Moreton}, G.~E. 1960, \aj, 65, 494

\bibitem[{{Muhr} {et~al.}(2011){Muhr}, {Veronig}, {Temmer}, {Kienreich}, \&
  {Vr{\v s}nak}}]{muhr11}
{Muhr}, N., {Veronig}, A.~M., {Temmer}, M., {Kienreich}, I., \& {Vr{\v s}nak},
  B. 2011, \apj, 739, 89

\bibitem[{{Patsourakos} \& {Vourlidas}(2009)}]{patsourakos09}
{Patsourakos}, S., \& {Vourlidas}, A. 2009, \apjl, 700, L182

\bibitem[{{Patsourakos} {et~al.}(2009){Patsourakos}, {Vourlidas}, {Wang},
  {Stenborg}, \& {Thernisien}}]{patsourakos09b}
{Patsourakos}, S., {Vourlidas}, A., {Wang}, Y.~M., {Stenborg}, G., \&
  {Thernisien}, A. 2009, \solphys, 259, 49

\bibitem[{{Temmer} {et~al.}(2011){Temmer}, {Veronig}, {Gopalswamy}, \&
  {Yashiro}}]{temmer11}
{Temmer}, M., {Veronig}, A.~M., {Gopalswamy}, N., \& {Yashiro}, S. 2011,
  \solphys, in press

\bibitem[{{Thompson} \& {Myers}(2009)}]{thompson09}
{Thompson}, B.~J., \& {Myers}, D.~C. 2009, \apjs, 183, 225

\bibitem[{{Thompson} {et~al.}(1998){Thompson}, {Plunkett}, {Gurman}, {Newmark},
  {St.~Cyr}, \& {Michels}}]{thompson98}
{Thompson}, B.~J., {Plunkett}, S.~P., {Gurman}, J.~B., {Newmark}, J.~S.,
  {St.~Cyr}, O.~C., \& {Michels}, D.~J. 1998, \grl, 25, 2465

\bibitem[{{Uchida}(1968)}]{uchida68}
{Uchida}, Y. 1968, \solphys, 4, 30

\bibitem[{{Veronig} {et~al.}(2010){Veronig}, {Muhr}, {Kienreich}, {Temmer}, \&
  {Vr{\v s}nak}}]{veronig10}
{Veronig}, A.~M., {Muhr}, N., {Kienreich}, I.~W., {Temmer}, M., \& {Vr{\v
  s}nak}, B. 2010, \apjl, 716, L57

\bibitem[{{Veronig} {et~al.}(2008){Veronig}, {Temmer}, \& {Vr{\v
  s}nak}}]{veronig08}
{Veronig}, A.~M., {Temmer}, M., \& {Vr{\v s}nak}, B. 2008, \apjl, 681, L113

\bibitem[{{Veronig} {et~al.}(2006){Veronig}, {Temmer}, {Vr{\v s}nak}, \&
  {Thalmann}}]{veronig06}
{Veronig}, A.~M., {Temmer}, M., {Vr{\v s}nak}, B., \& {Thalmann}, J.~K. 2006,
  \apj, 647, 1466

\bibitem[{{Wills-Davey} \& {Attrill}(2010)}]{wills10}
{Wills-Davey}, M.~J., \& {Attrill}, G.~D.~R. 2010, Space Science Reviews, 22

\bibitem[{{Zhukov}(2011)}]{zhukov11}
{Zhukov}, A.~N. 2011, Journal of Atmospheric and Solar-Terrestrial Physics, 73,
  1096

\end{thebibliography}

\end{document}